%% file: mainfile.tex
\documentclass[conference]{IEEEtran}
\usepackage{graphicx} 
\usepackage{color,soul}
\usepackage{paralist}
\usepackage{enumitem}
\usepackage{url}
\usepackage{amsfonts}
\usepackage{amssymb}
\usepackage{wrapfig}
\usepackage{amsmath}
\usepackage{subfigure}
\usepackage{gensymb}
\usepackage{multirow}
\usepackage{algorithm}
\usepackage{algorithmicx}
\usepackage{algpseudocode}
\usepackage{comment}
\usepackage{balance}
\usepackage[mathscr]{euscript}
\usepackage{caption}
\usepackage{float}
\usepackage{amsbsy}
\usepackage{bm}
\usepackage{amsthm}
\usepackage{amssymb}
\usepackage{diagbox}
\newcommand{\ignore}[1]{}

\theoremstyle{definition}

\renewcommand{\algorithmicrequire}{ \textbf{Input:}}

\title{Improving behavior based authentication against adversarial attack using XAI}

\author{
 \IEEEauthorblockN{Dong Qin, George Amariucai*, Daji Qiao, Yong Guan}
 \IEEEauthorblockA{
 {Iowa State University, 
 Ames, IA, USA},
 \{dqin, daji, guan\}@iastate.edu}
 \IEEEauthorblockA{{*Kansas State University, 
 Manhattan, KS, USA},
 amariucai@ksu.edu}\\
}

\begin{document}

\maketitle

\input{paper_body}

\bibliographystyle{IEEEtran}



\appendices

\input{supplementary_material}

\end{document}

%% file: paper_body.tex
\begin{abstract}
In recent years, machine learning models, especially deep neural networks, have been widely used for classification tasks in the security domain. However, these models have been shown to be vulnerable to adversarial manipulation: small changes learned by an adversarial attack model, when applied to the input, can cause significant changes in the output. Most research on adversarial attacks and corresponding defense methods focuses only on scenarios where adversarial samples are directly generated by the attack model. In this study, we explore a more practical scenario in behavior-based authentication, where adversarial samples are collected from the attacker. The generated adversarial samples from the model are replicated by attackers with a certain level of discrepancy. We propose an eXplainable AI (XAI) based defense strategy against adversarial attacks in such scenarios. A feature selector, trained with our method, can be used as a filter in front of the original authenticator. It filters out features that are more vulnerable to adversarial attacks or irrelevant to authentication, while retaining features that are more robust. Through comprehensive experiments, we demonstrate that our XAI based defense strategy is effective against adversarial attacks and outperforms other defense strategies, such as adversarial training and defensive distillation.

\end{abstract}


\input{introduction.tex}

\input{Problem_Statement}

\input{algorithm.tex}

\input{Comparison.tex}

\input{discussion.tex}

\input{related_work.tex}

%% file: introduction.tex
\section{Introduction}
\label{sec:introduction}

Machine learning models are rapidly gaining popularity in various critical domains of society, including the field of security. These models, particularly deep neural networks (DNNs), have been extensively applied in numerous scenarios, ranging from user recognition and intrusion detection to malware detection \cite{s20226578,2015DNS}.

However, it is important to note that there are potential risks associated with the misuse of these models. For instance, the 2024 Homeland Threat Assessment report of the US has highlighted that adversaries can exploit artificial intelligence to create more convincing misinformation and utilize this emerging technology to develop evasive cyberattacks on critical infrastructure. This emphasizes the need for robust security measures and continuous research to stay ahead of potential threats in this rapidly evolving landscape.

The adversarial use of deep learning models, known as adversarial attacks, has been extensively researched in recent years. These attacks involve manipulating input data, such as images, videos, or audio, to bypass deep learning-based authentication systems \cite{2020Adversarial, 2021Adversarial}. The goal of such attacks is to degrade the true positive rate of the detection model \cite{0The}. 

One challenge in defending against adversarial attacks is that the detection models are typically trained on a training set that follows a natural data distribution. However, the manipulated data used in adversarial attacks can be out of distribution (OOD) \cite{DBLP:journals/corr/abs-1811-01439}, leading to unexpected detection results in the model. 

Given the potential risks of adversarial manipulation attacks, it is crucial to understand how machine learning can be applied in security areas and how to mitigate these risks. Researchers and practitioners need to develop robust defenses and techniques to detect and counter adversarial attacks, ensuring the reliability and effectiveness of machine learning models in security applications.

Traditional user authentication methods, such as passwords, fingerprints, and retina scans, typically require stable and consistent inputs from valid users. Any deviation from the typical input, such as a single wrong digit in a password, leads to complete rejection. In contrast, behavioral biometrics, including gait parameters, hand gestures, and mouse operation dynamics, exhibit more variability and inconsistency. A user's distinctive behavioral pattern may occasionally deviate from their usual pattern. Therefore, behavioral authentication systems often need to collect a more substantial amount of behavioral data to ensure acceptable accuracy.

The nature of behavioral biometric data gives rise to a trait wherein if a subset of the input data closely matches the behavioral biometric authenticator, that specific input sample is more likely to be recognized as originating from the valid user. Unlike traditional user authenticators, which typically employ an AND logic to validate all inputs, behavioral user authenticators tend to utilize an OR logic for validation. However, this mechanism also renders behavioral authenticators more vulnerable to adversarial attacks, as attackers can manipulate only a portion of their input data to deceive the authenticator.

To mitigate the vulnerability of behavioral biometric authentication to adversarial attacks, we propose a novel defense method called XAI (explainable AI)-based defense strategy. This strategy is designed to enhance the robustness of the authenticator against such attacks. XAI has gained significant attention in recent years as it enables the interpretation of model classification results and enhances the transparency of the decision-making process \cite{lipton2018mythos, silver2017mastering, du2019techniques}. Within the field of XAI, feature attribution methods are widely used to explain model decisions on an instance-by-instance basis \cite{dabkowski2017real, chen2018learning, yoon2018invase, fu2021differentiated}. These methods provide insights into the relative importance of features for a given input data in determining the final classification.

Building upon the concept of feature attribution methods, we have developed a feature selector that is trained using the same training and testing set as the authenticator model. Our feature selector incorporates a state-of-the-art feature attribution method, enhanced with our novel improvements, to identify the key features that define the characteristics of a valid user. Simultaneously, it eliminates redundant features that have no relevance to the final decision-making process.

A crucial aspect of our feature selector is its ability to avoid selecting the unique behavioral patterns of valid users that exhibit significant fluctuations. This characteristic is advantageous from an attacker's perspective, as it becomes easier to mimic such patterns. By excluding these irrelevant or unstable features, our feature selector acts as a filter, effectively mitigating adversarial manipulations that specifically target these features.

In this research, our focus is on the application of the XAI-based feature selector within the context of behavioral biometric-based user authentication. In this scenario, the adversarial attacker can mimic the adversarial sample generated by an adversarial attack model. It is worth noting that unlike the image recognition scenario, the generated adversarial sample does not necessarily have to looks like the attacker's regular sample, it only needs to be recognized by the authenticator as a valid user's input.

\begin{figure*}[htb]
\centering\includegraphics[width=0.9\linewidth]{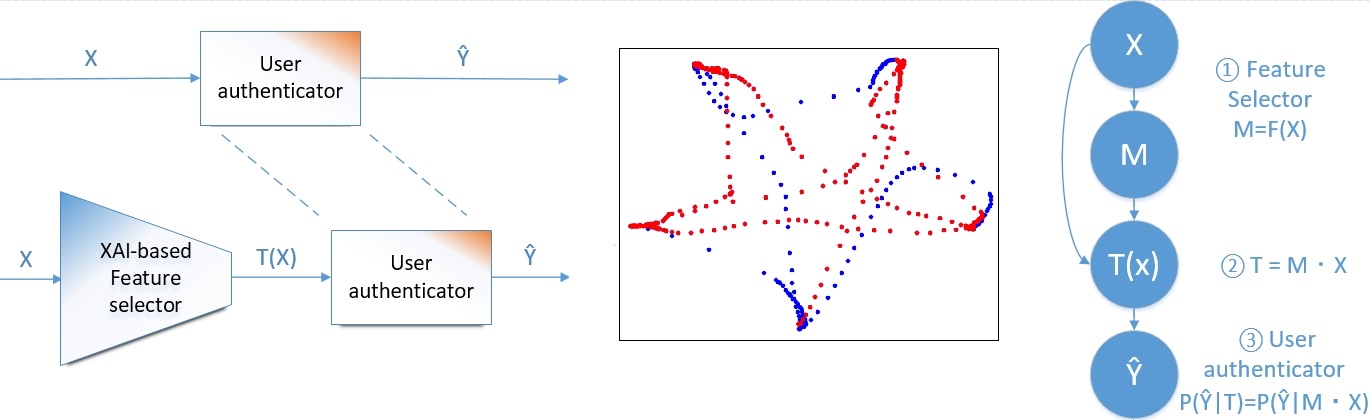}
\caption{Illustration of XAI augmented behavioral biometric authenticator. 4 mouse cursor movements (marked in blue) are selected as key part which is more important and robust than the rest 6 movements (marked in red) for user authentication. \textcircled{\small{1}} The feature selector takes an input x and returns a k-hot vector M which indicates whether each cognitive chunk, e.g. movement, will be selected as explanation for classification. \textcircled{\small{2}} $T=M\cdot X$ serves as a information bottleneck to get rid of the redundant and unreliable features. \textcircled{\small{3}} The user authenticator takes $T=M\cdot X$ for classification. In this research, feature selection means selecting part of the whole sequence rather than selecting a subset of all extracted dynamic features like velocity, acceleration, angular velocity and so on.}
\vspace{-4mm}
\label{fig:wd}
\end{figure*}

However, it is crucial to consider that attackers must possess the capability to physically execute the generated adversarial sample. In practice, there will inevitably be a certain level of discrepancy between the ideal adversarial sample and the actual attack sample created by the attacker. In other words, the real behavioral sample repeated by the attacker may not achieve as low true positive rate as the generated ideal sample can achieve. In such case, we demonstrate that our XAI-based feature selector can select key features that are less likely to be mimicked by attacker and can remove redundant features. Fig.~\ref{fig:wd} illustrates integrating our feature selector as a filter in front of the original authenticator. This integration significantly enhances the overall robustness of the classification system against adversarial attacks.

The contributions of this research can be summarized as follows:

1) We investigate the impact of adversarial attacks in a practical scenario in the context of behavioral biometric authentication, where the attacker has to accomplish the attack physically. As far as we know, we are the first one studying the adversarial attack in this scenario.

2) We propose a defense strategy based on feature attribution methods. By incorporating an adversarial attack-specific loss term during training, our feature selector can effectively identify features that are more resilient to adversarial attacks while still maintaining a good accuracy.

3) We compare our improved feature selector-based defense strategy with a basic feature selector-based defense strategy as well as adversarial training and defensive distillation based defense strategies using a mouse authentication dataset. The results clearly indicate that our proposed strategy outperforms all other defense strategies in terms of its effectiveness against adversarial attacks.

%% file: Problem_Statement.tex
\section{Problem Statement}
\label{sec:Problem Statement}

\subsection{Behavioral biometric authentication system}
In a behavioral biometric user authentication scenario, when a person attempts to log into a system, the user's behavioral data, such as mouse operations, hand gestures, and gait parameters, is collected by hardware devices like a mouse, camera, or inertial measurement unit. This data is then processed and fed into an authenticator model that has been trained using the valid user's previous behavioral data.

\begin{figure*}[htb]

\centering\includegraphics[width=1.5\columnwidth]{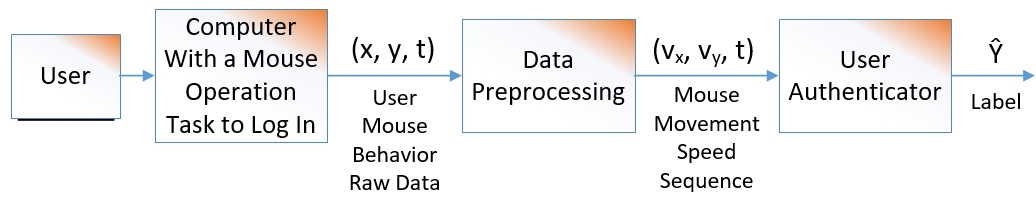}

\caption{A mouse behavior based user authentication system.}
\label{fig:authentication system}
\end{figure*}

Taking mouse-based user authentication as an example, as illustrated in Fig.~\ref{fig:authentication system}, when a user attempts to log in, their mouse movement trajectories during the login task is recorded as several movements in format of $(position, time)$ sequence. Subsequently, mouse dynamics such as speed and/or acceleration can be calculated from the raw input. The extracted mouse dynamics are then fed into a deep neural network-based authenticator. In our research, velocity sequence is used for classification due to its better overall performance.

\subsection{Threat model for behavioral biometric authentication}
\label{subsec:threat model}

In this paper, we focus on an attacking scenario where the authenticator is fully accessible to the attacker. However, in practical situations, attackers typically have limited access to the authenticator. Therefore, defense strategies that are effective against attacks with unlimited access should also be applicable to attacks with limited access to the authenticator.

In the case of a behavioral biometric authenticator, the adversarial attack is deployed in a more practical manner. In typical scenarios like image classification, the generated adversarial samples obtained from the adversarial model are directly fed into the classifier. However, for a behavioral biometric authenticator, the input data is collected directly from the person, such as gait parameters from a walking person or user-mouse interaction data from a PC/laptop user. As a result, the generated adversarial attack data has to be replicated by the attacker in practice in order to be effective. This adds additional complexity to the attack process.

To simulate the discrepancy between the generated ideal adversarial attack data and the real data collected from the attacker, we incorporate Gaussian noise into the generated ideal data. The calculation for determining the noise level is provided in Appendix~\ref{sec:noise level}. This level of discrepancy is determined based on the scenario described in the research presented by Mathew et al. \cite{mathew2019handedness}, where the attacker can manually control a cursor to follow the target cursor on the screen. Depending on the specific scenario, such as the absence of a targeting cursor or extensive prior practice, we can adjust the noise level accordingly to achieve a more accurate simulation.

In this research, we investigate two attack scenarios: one where the attacker has partial access to the detection system, and another where the attacker has complete access to the entire system.

\textbf{Attacking Scenario 1)} In the scenario of an adversarial attack where the authenticator is accessible, the attacker strategically deploys an attack against the authenticator. In order to achieve this, we train multiple attack models, each designed to generate a distinct movement that will replace the corresponding movement of the attacker's original sample. The ultimate goal is to minimize the true positive rate of the classifier. Subsequently, the attacker carefully selects the generated movement that results in the lowest true positive rate. As a result, the focus of the attack is narrowed down to effectively mimicking this specific movement during the login process, rather than targeting the entire sample as a whole.

\textbf{Attacking Scenario 2)} In addition to the scenario where we assume the attacker has access to the user authenticator model, we also explore an attacking scenario where the feature selector is also accessible. In this particular scenario, the attacker selectively chooses attack models that specifically target the movements identified by the feature selector as key features. As a result, the attacker's selection of the attack model that achieves the lowest true positive rate is made from a subset of attack models. This subset consists solely of attack models whose generated movement is guaranteed to be selected by the feature selector.

\begin{figure*}[htb]

\centering\includegraphics[width=1.5\columnwidth]{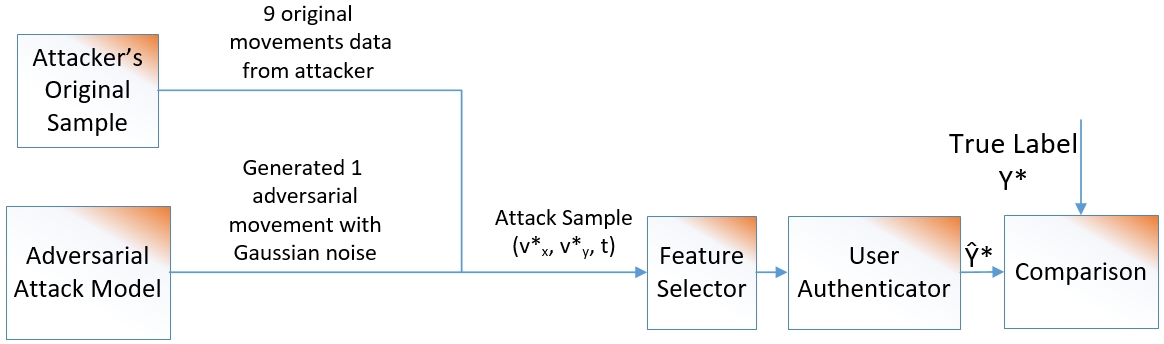}

\caption{In our experiment, each input sample includes 10 consecutive movements that follows certain pattern. 10 adversarial attack models are trained to generate 10 movements correspondingly to maximize prediction loss. 1 out of 10 (feature selector not accessible) or 1 out of 2/3/4/5 (feature selector accessible so attackers know which 2/3/4/5 movements are selected) models that achieves the lowest true positive rate is selected as the final attacking model.}
\label{fig:amtd}
\end{figure*}

Fig.~\ref{fig:amtd} illustrates the training diagram of the adversarial attack model. For simplicity and practicality, we only consider the situation where attacker can only mimic one adversarial movement during login. But the attacking scenario 2 can also be indicative of a situation where an attacker is capable of learning and replicating multiple movements during the login process. As the attacker tries to learn additional movements, the chances of those learned movements been selected by our feature selector increase. In attacking scenario 2, whether the attacker learns to mimic all movements or only the selected ones, it ensures that the adversarial movement will be selected and influence the final decision-making process. In such case, our feature selector contributes by forcing the attacker to mimic those movements selected by our feature selector that are more robust and harder to mimic. Consequently, the challenge to our feature selection based strategy changes to, whether focusing on several movements that are both crucial for classification and harder to mimic would improve the system overall performance compared to consider all movements even if there are some movements more vulnerable to adversarial attack.

In our research, the objective is to develop a defense strategy to protect against adversarial attacks. The performance of this defense strategy is primarily evaluated based on two key metrics. 1) True positive rate under adversarial attack, which measures the effectiveness of the authentication system in withstanding such attacks. This metric ensures the system's robustness and ability to accurately identify and reject adversarial samples. 2) Accuracy of the defense strategy in the normal testing set with valid and invalid user's data. This metric ensures that the defense strategy remains effective in practical everyday usage, guaranteeing the system's usefulness and reliability.

%% file: algorithm.tex
\section{The Proposed XAI based adversarial attack defense Strategy}
\label{sec:XAI based defense}

The feature selector in our research primarily utilizes the feature attribution method proposed in \cite{QIN2024106166} for training. However, in order to enhance the defense against adversarial attacks, our objective is to ensure that the feature selector not only filters out features with minimal contribution to classification accuracy but also filters out features that are particularly susceptible to adversarial attacks in practical scenarios. By incorporating this additional capability into the feature selector, we aim to further strengthen the system's resistance against adversarial attacks and improve its overall robustness.

\begin{figure*}[htb]

\centering\includegraphics[width=1.5\columnwidth]{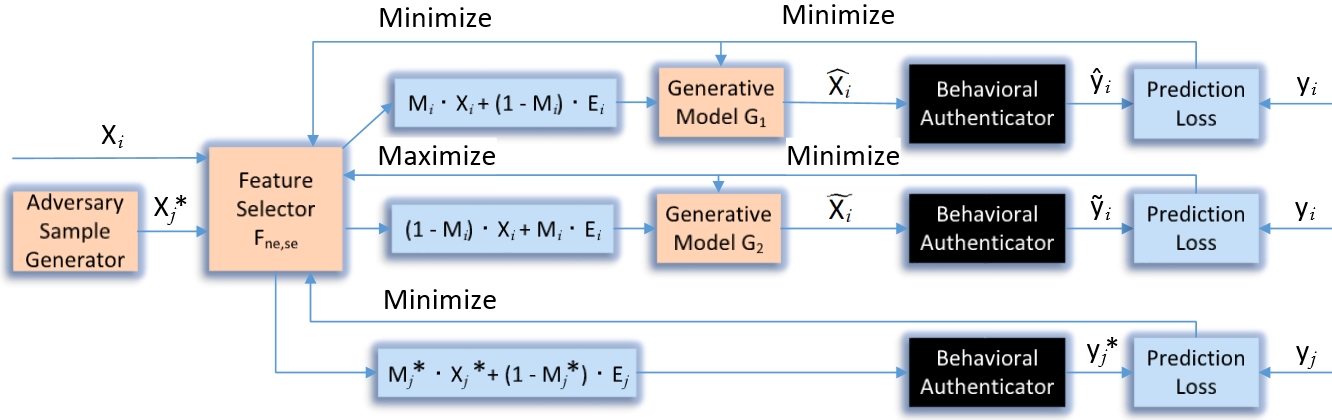}

\caption{Our improved feature selector training diagram.}
\label{fig:fstd}
\end{figure*}

In our research, as shown in Fig.~\ref{fig:fstd} we extend the training procedure proposed in the previous work \cite{QIN2024106166} by introducing an additional loop. This new loop aims to minimize the prediction loss when our classifier is under adversarial attack, where all features of the input adversarial sample $\bm{X}^*$ are generated by a well-trained adversarial sample generator. This adversarial sample generator generates adversarial samples with random Gaussian noise applied on each data point. This randomness simulates the natural discrepancy between the ideal adversarial samples and the real input samples collected from attackers.

By including this loss term in the training loop, our feature selector is encouraged to select features that are more difficult for attackers to mimic. In other words, the feature selector tends to prioritize features that represent valid user's stable and consistent patterns and discard features that are more specific to corner cases. For example, in Fig.~\ref{fig:boundary example}, the feature selector would likely to focus on feature X due to its smaller space between classifier boundaries for most valid user's samples. On the contrary, corner cases like the top left sample point will no longer be recognized as valid user's sample when we discard one feature, e.g. feature Y. Because feature Y of the classifier has enough space for discrepancy in that area. However, such sacrifice in true negative rate will either filter out the feature manipulated by adversarial model or force adversarial model to target on feature that has less tolerance for discrepancy.

In the following subsections, we will explain in detail how each part of our feature selector training diagram helps improve the behavioral authenticator system.

\subsection{XAI model selects key features}
 \label{subsec:XAI model}
A feature attribution method is applied to a machine learning classifier, generating a feature attribution vector for each input sample. This vector quantifies the contribution of each feature to the classification decision. Subsequently, a feature selector, based on the feature attribution vector, is employed to choose the top $n$ features for each input sample that have a greater impact on the classification outcome.

 In the user authentication scenario, redundant input features are commonly observed. In our research, we utilized a dataset of mouse-based user authentications, where users mainly rely on a few specific movements for effective classification. By employing a well-trained XAI algorithm-based feature selector, it is possible to identify the key movements of the whole input sample that significantly contribute to the classification process. By focusing on these essential movements, high classification accuracy can be achieved with less input features.

Therefore, to eliminate redundant features, we can train a feature selector using a state-of-the-art feature attribution algorithm \cite{QIN2024106166} and employ it as a filter. Fig.~\ref{fig:fstd} shows how this technique utilizes a generative model to handle common artifacts that may occur in conventional feature attribution methods. Moreover, the feature selector is optimized by minimizing the prediction loss associated with the selected features while maximizing the prediction loss associated with the non-selected features. This optimization approach effectively reduces the risk of information leakage through masking.

In our research, the preprocessed input data consists of sequences of $(v_x, v_y)$. For convenience, the feature selector will generate a vector indicating the selected movements rather than individual data points. We have conducted experiments, as described in Sec.\ref{sec:experiment}, to evaluate the performance of the feature selector and provide accurate results.

\subsection{Adversarial attack model achieves lower true positive rate when manipulating more vulnerable features}
\label{subsec:attacker model}

If the classifier is accessible, it is indeed possible to train an adversarial attack model that generates a portion of the input to replace the attacker's original input, resulting in a low true positive rate. However, in the case of a behavioral biometric data-based user authentication system, the authentication data is collected directly from a real person. Therefore, the adversarial sample generated by the attack model would need to be replicated by the attacker in the physical world. There will always be some level of discrepancy between the real behavioral data performed by a human and the target behavioral data generated by the adversarial attack model. As a result, the real sample played by a human may not be able to achieve as low of a true positive rate as the generated target sample.

In order to find those vulnerable features which the classifier boundaries in corresponding feature dimensions has larger tolerance for discrepancy and allows real world attackers to achieve low true positive rate, we add Gaussian noise to the adversary sample in the training of attack model. Therefore, by selecting one out of ten attack models that achieves the lowest true positive rate, we can find the most vulnerable one out of ten movements for the mouse authentication task that is easiest for attacker to mimic.

\begin{figure}[htb]

\centering\includegraphics[width=0.6\columnwidth]{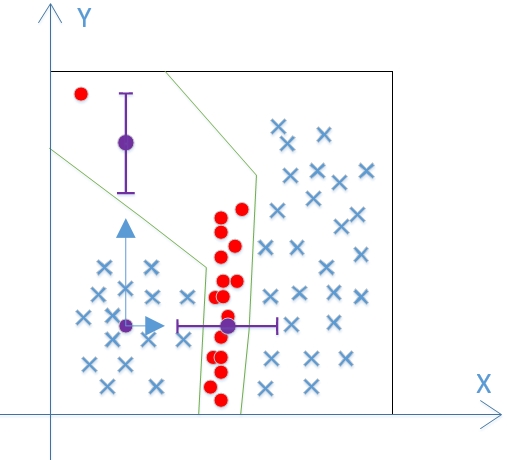}

\caption{After we apply random Gaussian noise (purple bars) to the generated feature dimension of the adversary sample (purple dots), more differentiable and consistent feature dimension (feature X) becomes more robust to the adversarial attack (part of the space between bars is out of classifier boundaries (green lines) of the valid user).}
\label{fig:boundary example}
\end{figure}

Fig.~\ref{fig:boundary example} provides an illustration in a two-dimensional example where the applied Gaussian noise improves the performance of the adversarial attack model by choosing to manipulate the feature that has a wider feature space between classifier boundaries. In this research, instead of selecting a single point, we select part of the entire velocity sequence, which we refer to as a movement for mouse behavior data. However, the analysis for low-dimensional cases can also be applied to higher-dimensional cases.

\subsection{Choosing important but less vulnerable features}
\label{subsec:discrepancy in features}

As shown in Fig.~\ref{fig:boundary example}, applying random Gaussian noise to the generated target features makes features that have a larger space between classifier boundaries more vulnerable to the adversarial attack. In contrast, features that are more consistent and stable, leading to a narrower feature space between classifier boundaries, are preferred for authentication system.

Therefore, our feature selector training diagram consists of two parts. The first part is the same as presented in \cite{QIN2024106166}, which allows the feature selector to choose features that minimize the prediction loss in the training set comprising data from both valid and invalid users. The second part involves a pre-trained adversary sample generator that targets the authenticator. By training the feature selector to minimize the prediction loss of adversarial samples in the authenticator, our feature selector can avoid selecting features that are vulnerable to adversarial manipulation, such as feature y in Fig.~\ref{fig:boundary example}. The trade-off between prediction loss in the regular training set and prediction loss in adversarial samples is controlled by the parameter $\beta$. The training process of our improved feature selector is summarized in Algorithm.\ref{alg:improved DoRaR}.

\begin{algorithm}

 \scriptsize
  \caption{Improved Feature Selector Training Algorithm}  
  \label{alg:improved DoRaR}  
    \begin{algorithmic}
    \Require
    $\bm{X}\in R^{s\times m}$, training set samples, where $s$ and $m$ are dataset size and feature size.
    $\bm{E}\in R^{s\times m}$, reference background noise used to filling the non-selected features, where $\bm{E}_i(j)$, is randomly sampled from feature $j$ in the training dataset.
    $\bm{X}_i'$, generated ideal adversarial sample.
    $\bm{N}_i(j) \sim \bm{N}(0,\sigma_i(j)^2)$, random Gaussian noise used to simulate the difference between generated ideal adversarial feature value and real feature value collected from attackers, where $\sigma_i(j)$ is determined by average velocity value of corresponding movement that $\bm{X}_i'(j)$ belongs to.
    $\bm{y}\in R^s$, ground truth labels for training set samples.
    $\bm{y}^*\in R^s$, ground truth labels for adversarial samples.
    $P$, behavioral authenticator;
    \Ensure  
     $F_{n_e,s_e}(\cdot;\theta_F)$, feature selector that returns mask to select $n_e$ chunks each consist of $s_e$ features given an input instance $\bm{X}$
    \renewcommand{\algorithmicrequire}{ \textbf{Select:}}
    \Require
    $d$, mini-batch size;
    $\lambda$, learning rate;
    $t$, training steps;
    $\alpha$, control parameter that balance selected and non-selected features;
    $\beta$, control parameter that balance prediction loss from training set samples and adversarial samples;
    $n_e$, number of output explanation unit;
    $s_e$, size of output explanation unit;
    \For{$1,...,t$}
        \State Randomly sample mini-batch of size $d$, $\bm{X}\in R^{d\times m}$

        \For{$i = 1,...,d$}
            \State
            $\bm{M}_i=F(\bm{X}_i;\theta_F)$
            \State $\widehat{\bm{X}}_i=G_1(\bm{X}_i\cdot \bm{M}_i+\bm{E}_i\cdot(1-\bm{M}_i);\theta_{G_1})$
            \State $\tilde{\bm{X}_i}=G_2(\bm{X}_i\cdot (1-\bm{M}_i)+\bm{E}_i\cdot \bm{M}_i;\theta_{G_2})$

            \State            $\widehat{\bm{y}}_i=P(\widehat{\bm{X}}_i)$
            \State
            $\tilde{\bm{y}_i}=P(\Tilde{\bm{X}_i})$

            \State
            $\bm{X}^*_i = \bm{X}_i' + \bm{N}_i$
            \State
            $\bm{M}^*_i=F(\bm{X}^*_i;\theta_F)$

            \State
            $\widehat{\bm{y}}_i^*=P(\bm{X}^*_i\cdot \bm{M}^*_i+\bm{E}_i\cdot(1-\bm{M}^*_i))$
            
            \State
            $l_1=-\frac{\sum_{i=1}^{d}\bm{y}_i\log(\widehat{\bm{y}}_i)}{d}$
            \State
            $l_2=-\frac{\sum_{i=1}^{d}\bm{y}_i\log(\tilde{\bm{y}}_i)}{d}$
            
            \State
            $l_3=-\frac{\sum_{i=1}^{d}\bm{y}^*_i\log(\widehat{\bm{y}}_i^*)}{d}$
        \EndFor
        \State
        $\theta_F = Adam(L=(1-\alpha)l_1-\alpha l_2 + \beta l_3,\lambda)$
        \State
        $\theta_{G_1} = Adam(L=l_1,\lambda)$
        \State
        $\theta_{G_2} = Adam(L=l_2,\lambda)$
    \EndFor

    \end{algorithmic}

\end{algorithm}

%% file: Comparison.tex
\section{Experiment on mouse based user authentication dataset}
\label{sec:experiment}

\subsection{Evaluated adversarial attack defense strategies}
In our experiment, we compare four different adversarial attack defense strategies: 

1) The \textbf{Improved feature selector based defense strategy}, as illustrated in Fig.~\ref{fig:fstd} and Algorithm~\ref{alg:improved DoRaR}. 

2) The \textbf{Basic feature selector based defense strategy}, which is similar to the first strategy, but without including an adversarial sample generator in the training loop. 

For both feature selector based defense methods, we experimented with four different feature selectors. These feature selectors are trained to select 2, 3, 4, or 5 movements.

3) The \textbf{Adversarial training based defense strategy}, which involves adding adversarial samples generated by an adversarial sample generator to the training set. This new training set is then used to retrain a new behavior authenticator. The adversarial sample generator used in this strategy has the same structure as the one employed by attackers.

4) The \textbf{Defensive distillation based defense strategy}, which takes advantageous of the model distillation method. It uses the output score of the regular model as the soft label to train the distilled model. The distilled model used in defensive distillation has the same structure as the regular model but is less sensitive to small perturbation.

Further details regarding the model structure and parameter settings can be found in Appendix.~\ref{sec:model structure}.

\subsection{Mouse behavior based user authentication dataset}
The dataset used to evaluate all four adversarial defense methods is the mouse behavior authentication dataset presented in \cite{fu2022artificial}. In this dataset, 21 subjects were instructed to perform a predefined task consisting of 10 consecutive movements in a specific pattern, as shown in Fig.~\ref{fig:wd}. Each movement has a defined starting and ending point.

The raw data is collected in the form of $(x, y, t)$, where $x$ and $y$ represent the coordinates of the mouse cursor, and $t$ represents the time stamp. However, for classification purposes, the velocity $(v_x, v_y)$ is calculated from the raw data and used as input features. The decision to use velocity instead of raw data is based on its better overall performance in classification tasks.

\begin{figure}[t]

\centering\includegraphics[width=0.6\columnwidth]{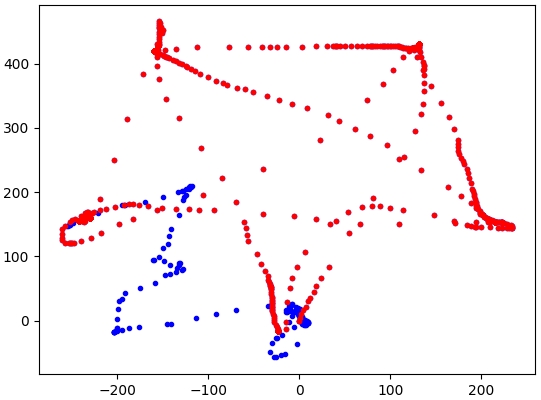}

\caption{Example of authentication task played by an attacker: velocity sequence is transferred to position domain for better illustration, attacker's sample with 1 movement (marked in blue) replaced by generated adversarial movement with applied Gaussian noise.}
\label{fig:adv_sample}
\end{figure}

The dataset consists of 21 subjects, varying in age, gender, and education background, all of whom possess proficiency in using a mouse. Each subject's data comprises a minimum of 66 trials. From these subjects, 18 are selected for training 18 behavior authenticators, with one subject serving as the valid user and the remaining 17 as invalid users for training and testing purposes. The last three subjects are designated as adversarial attackers, where one out of ten movements in their samples is replaced with adversarial sample data. Fig.~\ref{fig:adv_sample} shows an example of the adversarial attack sample.

\subsection{Experiment Result}

\begin{table*}[hbt]
\centering
\caption{Statistical comparison for improved feature selector based defense strategy and three strategies been compared.}
\label{table:Comparison results}
\scalebox{0.67}{
\begin{tabular}{|cc|cccccccccccclll|}
\hline
\multicolumn{2}{|c|}{Our defense strategy} & \multicolumn{15}{c|}{\begin{tabular}[c]{@{}c@{}}Improved feature selector based defense strategy\\ (shows in the left side of the inequality sign)\end{tabular}} \\ \hline
\multicolumn{2}{|c|}{\begin{tabular}[c]{@{}c@{}}Defense strategy\\ been compared\end{tabular}} & \multicolumn{4}{c|}{\begin{tabular}[c]{@{}c@{}}Adversarial training based\\ defense strategy\end{tabular}} & \multicolumn{4}{c|}{\begin{tabular}[c]{@{}c@{}}Defensive distillation\\ based defense strategy\end{tabular}} & \multicolumn{7}{c|}{\begin{tabular}[c]{@{}c@{}}Basic feature selector\\ based defense strategy\end{tabular}} \\ \hline
\multicolumn{2}{|c|}{\begin{tabular}[c]{@{}c@{}}Number of movements\\ been selected\end{tabular}} & \multicolumn{1}{c|}{2} & \multicolumn{1}{c|}{3} & \multicolumn{1}{c|}{4} & \multicolumn{1}{c|}{5} & \multicolumn{1}{c|}{2} & \multicolumn{1}{c|}{3} & \multicolumn{1}{c|}{4} & \multicolumn{1}{c|}{5} & \multicolumn{1}{c|}{2} & \multicolumn{1}{c|}{3} & \multicolumn{1}{c|}{4} & \multicolumn{4}{c|}{5} \\ \hline
\multicolumn{2}{|c|}{\begin{tabular}[c]{@{}c@{}}Accuracy in\\ testing set\end{tabular}} & \multicolumn{1}{c|}{\begin{tabular}[c]{@{}c@{}}0.702\textless{}0.876\\ p=1.992e-07\end{tabular}} & \multicolumn{1}{c|}{\begin{tabular}[c]{@{}c@{}}0.751\textless{}0.876\\ p=2.372e-05\end{tabular}} & \multicolumn{1}{c|}{\begin{tabular}[c]{@{}c@{}}0.809\textless{}0.876\\ p=5.092e-04\end{tabular}} & \multicolumn{1}{c|}{\begin{tabular}[c]{@{}c@{}}0.851\textless{}0.876\\ p=0.106\end{tabular}} & \multicolumn{1}{c|}{\begin{tabular}[c]{@{}c@{}}0.702\textless{}0.896\\ p=1.260e-08\end{tabular}} & \multicolumn{1}{c|}{\begin{tabular}[c]{@{}c@{}}0.751\textless{}0.896\\ p=9.243e-07\end{tabular}} & \multicolumn{1}{c|}{\begin{tabular}[c]{@{}c@{}}0.809\textless{}0.896\\ p=1.749e-05\end{tabular}} & \multicolumn{1}{c|}{\begin{tabular}[c]{@{}c@{}}0.851\textless{}0.896\\ p=0.005\end{tabular}} & \multicolumn{1}{c|}{\begin{tabular}[c]{@{}c@{}}0.702\textless{}0.717\\ p=0.161\end{tabular}} & \multicolumn{1}{c|}{\begin{tabular}[c]{@{}c@{}}0.751\textless{}0.764\\ p=0.256\end{tabular}} & \multicolumn{1}{c|}{\begin{tabular}[c]{@{}c@{}}0.809\textgreater{}0.805\\ p=0.385\end{tabular}} & \multicolumn{4}{c|}{\begin{tabular}[c]{@{}c@{}}0.851\textless{}0.852\\ p=0.468\end{tabular}} \\ \hline
\multicolumn{1}{|c|}{\multirow{2}{*}{\begin{tabular}[c]{@{}c@{}}True positive\\ rate under\\ adversarial\\ attack\end{tabular}}} & \begin{tabular}[c]{@{}c@{}}Classifier\\ accessible\end{tabular} & \multicolumn{1}{c|}{\begin{tabular}[c]{@{}c@{}}0.724\textgreater{}0.195\\ p=1.266e-04\end{tabular}} & \multicolumn{1}{c|}{\begin{tabular}[c]{@{}c@{}}0.674\textgreater{}0.195\\ p=0.001\end{tabular}} & \multicolumn{1}{c|}{\begin{tabular}[c]{@{}c@{}}0.658\textgreater{}0.195\\ p=0.001\end{tabular}} & \multicolumn{1}{c|}{\begin{tabular}[c]{@{}c@{}}0.463\textgreater{}0.195\\ p=0.040\end{tabular}} & \multicolumn{1}{c|}{\begin{tabular}[c]{@{}c@{}}0.724\textgreater{}0.118\\ p=9.750e-07\end{tabular}} & \multicolumn{1}{c|}{\begin{tabular}[c]{@{}c@{}}0.674\textgreater{}0.118\\ p=5.431e-06\end{tabular}} & \multicolumn{1}{c|}{\begin{tabular}[c]{@{}c@{}}0.658\textgreater{}0.118\\ p=4.768e-06\end{tabular}} & \multicolumn{1}{c|}{\begin{tabular}[c]{@{}c@{}}0.463\textgreater{}0.118\\ p=0.005\end{tabular}} & \multicolumn{1}{c|}{\begin{tabular}[c]{@{}c@{}}0.724\textgreater{}0.644\\ p=0.090\end{tabular}} & \multicolumn{1}{c|}{\begin{tabular}[c]{@{}c@{}}0.674\textgreater{}0.614\\ p=0.086\end{tabular}} & \multicolumn{1}{c|}{\begin{tabular}[c]{@{}c@{}}0.658\textgreater{}0.569\\ p=0.070\end{tabular}} & \multicolumn{4}{c|}{\begin{tabular}[c]{@{}c@{}}0.463\textgreater{}0.388\\ p=0.094\end{tabular}} \\ \cline{2-17} 
\multicolumn{1}{|c|}{} & \begin{tabular}[c]{@{}c@{}}Classifier and\\ feature\\ selector\\ accessible\end{tabular} & \multicolumn{1}{c|}{\begin{tabular}[c]{@{}c@{}}0.384\textgreater{}0.195\\ p=0.064\end{tabular}} & \multicolumn{1}{c|}{\begin{tabular}[c]{@{}c@{}}0.316\textgreater{}0.195\\ p=0.144\end{tabular}} & \multicolumn{1}{c|}{\begin{tabular}[c]{@{}c@{}}0.303\textgreater{}0.195\\ p=0.160\end{tabular}} & \multicolumn{1}{c|}{\begin{tabular}[c]{@{}c@{}}0.284\textgreater{}0.195\\ p=0.243\end{tabular}} & \multicolumn{1}{c|}{\begin{tabular}[c]{@{}c@{}}0.384\textgreater{}0.118\\ p=0.006\end{tabular}} & \multicolumn{1}{c|}{\begin{tabular}[c]{@{}c@{}}0.316\textgreater{}0.118\\ p=0.016\end{tabular}} & \multicolumn{1}{c|}{\begin{tabular}[c]{@{}c@{}}0.303\textgreater{}0.118\\ p=0.018\end{tabular}} & \multicolumn{1}{c|}{\begin{tabular}[c]{@{}c@{}}0.284\textgreater{}0.118\\ p=0.040\end{tabular}} & \multicolumn{1}{c|}{\begin{tabular}[c]{@{}c@{}}0.384\textgreater{}0.317\\ p=0.029\end{tabular}} & \multicolumn{1}{c|}{\begin{tabular}[c]{@{}c@{}}0.316\textless{}0.327\\ p=0.279\end{tabular}} & \multicolumn{1}{c|}{\begin{tabular}[c]{@{}c@{}}0.303\textgreater{}0.285\\ p=0.176\end{tabular}} & \multicolumn{4}{c|}{\begin{tabular}[c]{@{}c@{}}0.284\textgreater{}0.245\\ p=0.074\end{tabular}} \\ \hline
\end{tabular}
}
\vspace{-2mm}
\end{table*}

\begin{figure*}[htb]
\centering\includegraphics[width=2.1\columnwidth]{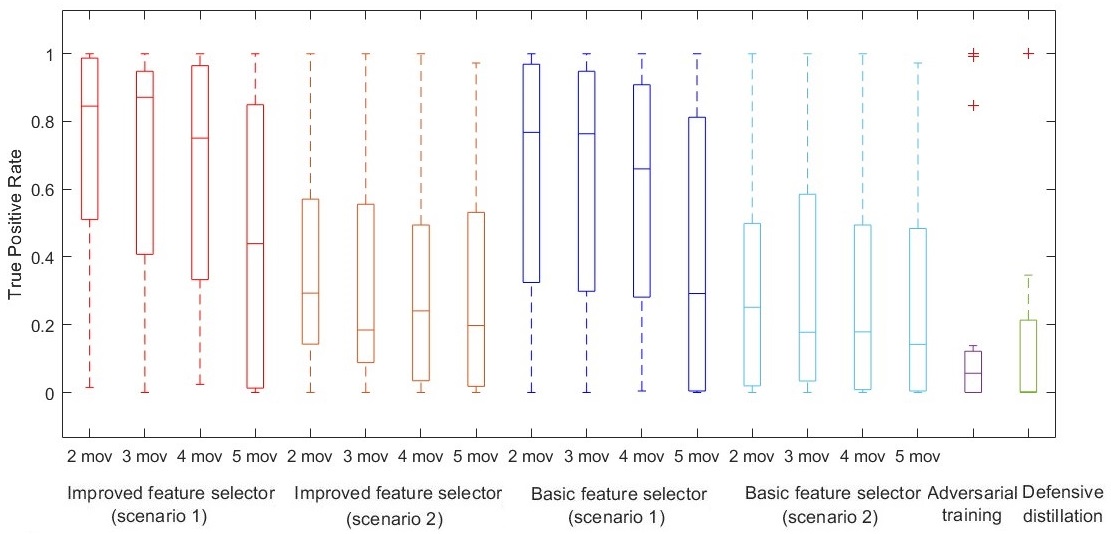}
\caption{True positive rate of 4 defense strategies under adversarial attack.}
\label{fig:TPR}
\end{figure*}

\begin{figure*}[htb]
\centering\includegraphics[width=2\columnwidth]{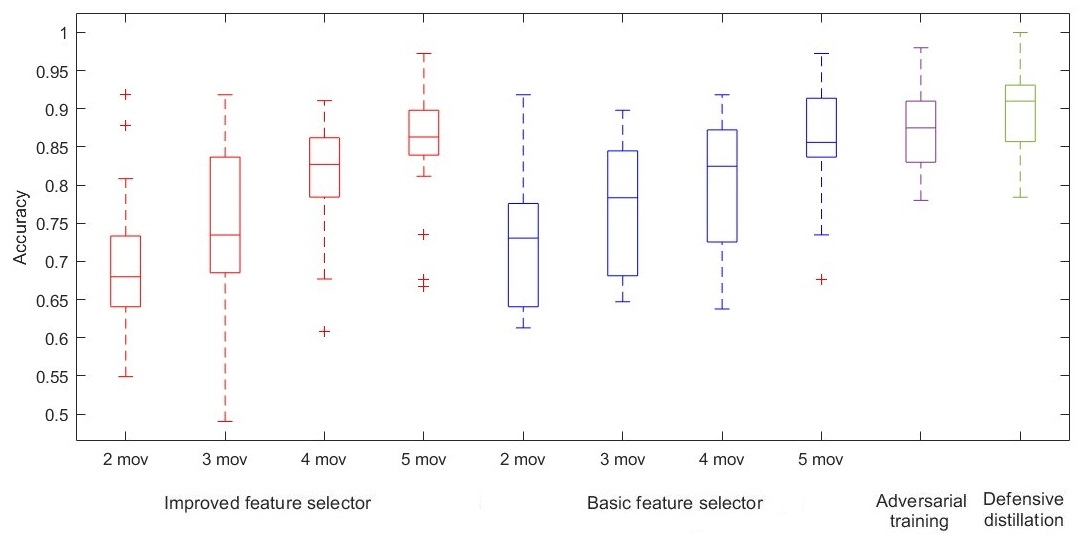}
\caption{Accuracy of 4 defense strategies in normal testing test.}
\label{fig:ACC}
\end{figure*}

We tested two feature selector based defense strategies in two attacking scenarios as introduced in section.~\ref{subsec:threat model}. For each feature selector based strategy, we tested 4 feature selectors to select 2,3,4 and 5 movements as well as keeping all 10 movements. For the adversarial training and defensive distillation based strategies, since there is no feature selector, we simply get the corresponding true positive rate and accuracy for 18 authenticators.

Fig.~\ref{fig:TPR} and Fig.~\ref{fig:ACC} summarizes the overall result in adversarial sample set and normal testing set. Table.~\ref{table:Comparison results} show the one-tailed paired t-test comparison result between our strategy and three strategies been compared, where result of our strategy is shown in the left side of inequality sign and other three strategies are shown in the right side.

The threshold for authenticators in both testing set and adversarial sample set are set to the default value as it is in the training set. For the detailed experiment results, see Table.~\ref{table:improved feature selector} to Table.~\ref{table:basic feature selector ts} in Appendix.~\ref{sec:detailed results}.

\subsubsection{Performance under adversarial attack}
\label{subsubsec:Performance under adversarial attack}

As shown in Table.~\ref{table:Comparison results} and Fig.~\ref{fig:TPR}. Our method demonstrates a significant improvement in true positive rate against adversarial attacks compared to the adversarial training and defensive distillation based strategy. Only when both classifier and feature selector are accessible, comparing to the adversarial training based strategy, the improvement is insignificant when selecting 3,4 or 5 movements for the significance level equal to $\alpha = 0.1$.

Additionally, comparing to the basic feature selector. Our strategy outperforms that defense strategy under adversarial attacks in the scenario when only classifier is accessible. When both classifier and feature selector are accessible, for selecting 3 or 4 movements, there is no significant improvement from our strategy.

The improvement from our method can be attributed to the improved feature selector's ability to avoid selecting movements that are vulnerable to adversarial attacks. In other words, the feature selector discards features that correspond to classifier boundaries with a large tolerance for discrepancies between generated ideal adversarial data and real data, even if these features may enhance classification accuracy.

There are two trends we can see from the results. First, for attacking scenario 2 when both classifier and feature selector are accessible from attackers, the improvement from our method becomes less comparing to attacking scenario 1. Second, when the number of selected movements increases, the improvement from our method becomes less.

However, even in the worst-case scenario where the feature selector is accessible and it selects 5 movements for classification, our method still demonstrates a improvement in the true positive rate for adversarial samples. Specifically, our method improves the true positive rate by 16.3\%, 45.6\% and 140.7\% compared to the basic feature selector, adversarial training and defensive distillation respectively. But such improvement is not evenly distributed in all 18 authenticators. On the contrary, it improve certain authenticators a lot while for other authenticators, the improvement from our method is less. More specifically, the more overlap there is between important features and invulnerable features for a authenticator, the more robustness our method can bring to it against adversarial attacks. This explains the relatively high $p$-value in some scenarios.

\subsubsection{Performance in normal testing set}
\label{subsubsec:Performance in normal testing set}

Table.~\ref{table:Comparison results} and Fig.~\ref{fig:ACC} shows the classification accuracy in the testing set. When comparing to the adversarial training-based authenticator, the classification accuracy of our strategy decreases by 19.5\% to 2.4\% when the feature selector selects 2 to 5 movements. When comparing to the defensive distillation-based authenticator, accuracy of our method decreases by 21.7\% to 5.0\% when selecting 2 to 5 movements. Comparing to the basic feature selector-based strategy, the accuracy in the testing set is lower but not significant. Considering the significant improvement in true positive rate, our method strikes a balance between maintaining classification accuracy and improving defense against adversarial attacks.

\subsubsection{ROC curves for valid user and adversarial attackers}
\label{subsubsec:ROC curves}

\begin{figure*}[h]
\centering
    \subfigure[Input data: valid user's samples and attackers' samples without adversarial modification. Feature selector selecting 2, 3, 4, 5 movements.]{\label{fig:guided0}\includegraphics[width=1\columnwidth]{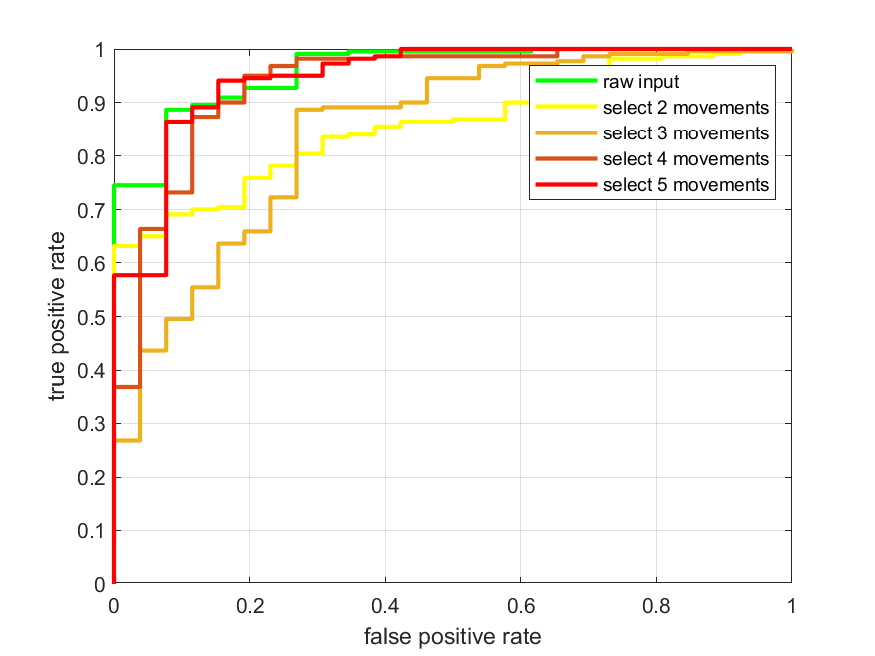}}
\centering
    \subfigure[Input data: valid user's samples and attackers' samples with adversarial modification. Feature selector selecting 2, 3, 4, 5 movements.]{\label{fig:Ours0}\includegraphics[width=1\columnwidth]{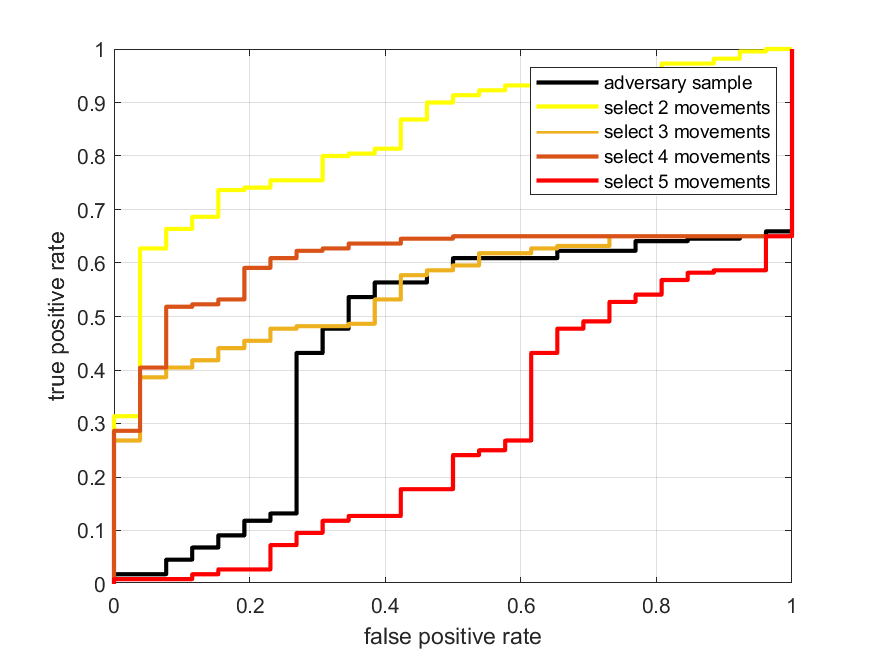}}
\centering
\caption{Example of ROC curves for given input data from one valid user and three unknown attackers.}
\vspace{-1mm}
\label{fig:ROC}
\end{figure*}

To provide a clearer illustration, we have plotted a set of ROC curves in Fig.~\ref{fig:ROC}, which combines data from valid users and adversarial attackers. By examining these ROC curves and referring to the results from previous tables, we can observe a trend: the more movements we select, the higher accuracy we can achieve in the common testing set when there is no adversarial modification.

When adversarial modification occurs, the performance of the authenticator that takes all 10 movements as input shows a significant decrease. On the other hand, the feature selector-based authentication system demonstrates superior performance, particularly when the selected movements do not include the modified movement. For instance, if the feature selector chooses 2 movements, the system achieves higher true positive rate. However, as the number of selected movements by the feature selector increases to 5, the overall true positive rate quickly drops due to the inclusion of the modified movement within the selection. Based on our observations from all experiments, we have noticed a consistent trend: the more movements our feature selector selects, the higher the likelihood of including the modified movement in its selection.

Nevertheless, even if the feature selector is accessible and the modified movement falls within the selected movement range, the overall performance remains superior compared to the scenario without a feature selector. This is because the presence of a feature selector compels the attacker to mimic a less vulnerable movement, which has a lower tolerance for discrepancies between generated ideal adversarial data and real data. Consequently, our method enhances the robustness of the authentication system against adversarial attacks, regardless of the attacker's ability to mimic multiple movements. The performance upper bound of our method is actually depend on the user's stable and consistent mouse behavior pattern. Although corner cases, such as the top left sample in Fig.~\ref{fig:boundary example}, can be learned by the classifier, they are disregarded by the feature selector-classifier combination system due to its vulnerable key feature pattern.

%% file: discussion.tex
\section{Conclusion}
\label{sec:Conclusion}

In this paper, our investigation focuses on the potential application of explainable AI (XAI) in defending against adversarial attacks in the context of behavioral biometric authentication. We specifically explore a category of XAI known as feature attribution methods, which are utilized to train a feature selector responsible for selecting key features and eliminating redundant ones. In our study, we conduct tests on two types of feature selectors: the basic feature selector used in XAI research \cite{QIN2024106166}, and our improved version that incorporates an adversarial sample generator within the training loop. We consider two attacking scenarios: one where the authenticator is fully accessible, and another where both the authenticator and feature selector are accessible to the attacker. Additionally, we compare our approach to other commonly used defense strategies against adversarial attacks, specifically adversarial training-based defense method and defensive distillation-based defense method.

To evaluate the aforementioned defense methods, we implemented them using a mouse behavior-based user authentication dataset introduced in \cite{fu2022artificial}. Through comprehensive testing, several conclusions can be drawn. Firstly, incorporating a feature selector in the authentication system significantly enhances its security against adversarial attacks. Secondly, when an adversarial sample generator is included in the training loop of the feature selector, the system's performance against adversarial attacks further improves. Thirdly, the number of features selected by the feature selector directly impacts the classification accuracy in common data sets and true positive rate under adversarial attack.

Indeed, while there may be a slight decrease in classification accuracy in the common testing set due to the feature selector, this decrease can be limited to a small range if the valid user exhibits several stable and consistent behavior patterns, and an appropriate number of features is selected, e.g. selecting 5 movements. On the other hand, the addition of the improved feature selector leads to a significant enhancement in the authentication system's resilience against adversarial attacks. Therefore, the benefits of adding the feature selector outweigh the disadvantage of slight decrease in classification accuracy.

%% file: related_work.tex
\section{Related Work}
\label{subsec:related-work}

Various defense mechanisms have been extensively studied to protect detection systems against adversarial attacks. One such method, known as adversarial training \cite{DBLP:journals/corr/GoodfellowSS14}\cite{7780651}\cite{2013Intriguing}, involves augmenting the training dataset with adversarial samples. This technique incorporates adversarial samples generated by one or more adversarial attack models into the training set, followed by retraining the classifier. Although adversarial training improves the classifier's robustness in scenarios like image recognition, its effectiveness may be limited in different scenarios such as behavioral biometric authentication. In this specific scenario, the generated adversarial samples are allowed to closely resemble the valid user's sample, potentially leading to a significant reduction in the model's true negative rate if added to the training set. Thus, alternative defense mechanisms need to be explored to ensure the security of behavioral biometric authentication systems.

Defensive distillation \cite{2016Distillation} is a defense mechanism that trains the classifier in two rounds using a modified version of the distillation method \cite{2015Distilling}. This approach has the advantage of producing a smoother network and decreasing the magnitude of gradients around input points. As a result, it becomes more challenging for attackers to generate adversarial examples \cite{2016Distillation}. However, studies have revealed that while defensive distillation proves effective against white-box attacks, it falls short in providing sufficient protection against black-box attacks that are transferred from other networks \cite{2016Towards}. Therefore, further research is necessary to develop defense mechanisms that can effectively counter black-box attacks and enhance the security of detection systems.

The feature selection methods mentioned in \cite{2009Evade,Battista2013Security, 2014Feature, Fei2016Adversarial, 8249883} are the most similar approaches to our strategy. These studies investigated the use of reduced feature sets for classification as a defense against adversarial attacks. However, there are two significant differences between these approaches and our research. 

First, these defense methods are typically tested against pure cyberattacks where the input data is directly collected from the adversarial attack model. However, in the case of behavioral biometric authentication, where the input data is collected from humans, there are limitations on the attacker's ability to accurately mimic the generated adversarial sample. Besides, unlike in scenarios like image recognition, where the modified adversarial sample must resemble the original sample from a human perspective, the modified attacker's behavioral biometric sample does not necessarily need to be similar to the original attacker's sample.

Second, it is important to note that the feature selection methods discussed in \cite{2009Evade, Battista2013Security, 2014Feature, Fei2016Adversarial, 8249883} are not instance-wised feature selection, they discard the same features regardless of the input. In contrast, our XAI-based feature selector achieves sample-specific feature selection. As a result, our feature selector can ensure a higher level of classification accuracy.

%% file: supplementary_material.tex
\section{Gaussian noise level to simulate difference between ideal adversarial sample and real sample}
\label{sec:noise level}
In order to simulate the difference between ideal adversarial mouse behavior data and real data collected from attackers in the physical world, we use the result from \cite{mathew2019handedness}. We assume the relative position of the manually controlled cursor in x or y coordinate regarding to the target cursor follows a random Gaussian distribution, so the expected distance between manually controlled cursor and target cursor in x or y coordinate equals to
\begin{equation}
       E(d)=\int_{-\infty }^{\infty }\left| x\right| \frac{1}{\sqrt{2\pi}\sigma_p }e^{-\left ( \frac{x^2}{2\sigma_p^2} \right )}dx=\sqrt{\frac{2}{\pi}}\sigma_p\approx 0.8\sigma_p
\end{equation}
Therefore, the average distance between the manually controlled cursor and the target cursor in x or y coordinate of the manipulated movement with length 160 equals to
\begin{equation}
    \overline{d}=0.8\overline{\sigma_p}=0.8\frac{\sum_{i=1}^{160}\sigma_{p_i}}{160}
\end{equation}
In each point of the velocity sequence there is an applied Gaussian perturbation to the generated target velocity value, so the accumulated position error follows a Gaussian distribution with $\sigma$ equals to:
\begin{equation}
    \sigma_{p_i}^{2}=\sum_{n=1}^{i}\sigma_{v_n}^{2}
\end{equation}
where $\sigma_{p_i}$ is sigma of the accumulated Gaussian error of $i$-th point in position domain, and $\sigma_{v_n}$ is sigma of the applied Gaussian perturbation of $n$-th point in velocity domain.

Assuming we apply a constant Gaussian perturbation which follows $N(0,\sigma^2)$ to each point of velocity sequence in both x and y coordinate, s.t.
\begin{equation}
    \sigma_{p_i}=\sqrt{i}\cdot\sigma
\end{equation}
Therefore, we have
\begin{equation}
    \overline{d}=0.8\frac{\sum_{i=1}^{160}\sigma_{p_i}}{160} = \frac{0.8}{160}\sigma*\sum_{i=1}^{160}\sqrt{i} \approx 6.78\sigma
\end{equation}
From \cite{mathew2019handedness}, we have the ratio of the average cursor speed to the average distance between target cursor and manually controlled cursor,
\begin{equation}
    \overline{d}=\frac{\overline{v}}{8}
\end{equation}
Therefore,
\begin{equation}
    \sigma=0.0184\overline{v}
\end{equation}
the standard deviation of applied Gaussian noise to the velocity sequence can be calculated from the average value of that sequence.

\section{Model structure of feature selector, generative model, behavior authenticator and adversarial sample generator}
\label{sec:model structure}
In our experiments, for the feature selector, it has 1 2D convolutional layer with kernel size $2\times9$ and 2 1D convolutional layers with kernel size 7 and 5. These 3 CNN layers have 64, 96 and 1 filters respectively. First two CNN layers are followed by Max pooling layer with pool size 2 and Batch norm layer. Then there is a LSTM network with 2 recurrent layers. After that, there are two fully connected layers with 400 and 10 output units. 10 output scores correspond to 10 movement of 1 mouse based authentication task. Relu is chosen as the active function. Learning rate for feature selector is set to $\lambda = 5e-4$. $\alpha$ is set to the ratio of selected movements to all movements which is 10 in our experiment. $\beta$ is set to the value that guarantees a close accuracy $(\leq 95\%)$ to the accuracy achieved by basic feature selector (without adversarial sample generator in the training loop) in the validation set, under this requirement, the parameter value that achieves the highest true positive rate in adversarial samples is selected.

According to the sample size of the mouse dataset, the generative model consists of 2 fully connected linear layers both with 3200 units. Learning rate for generative models are set to $\lambda = 1e-4$.

The behavior authenticator model in this research has 3 CNN layers each with a dropout layer in the front. First and second CNN layer are followed by a batch norm layer. 3 CNN layers have 64, 96, 128 kernels with kernel size (2,9), 7 and 5. After that, there is LSTM network with 128 input size, hidden size and 2 layers. At the end, there are two fully connected layers with 128 and 2 output units. Relu is chosen as the active function for CNN and fully connected layers. Learning rate for behavior authenticator is set to $\lambda = 1e-4$.

The adversarial sample generator model consists of 2 fully connected layers each with 320 output units, so it can generate velocity sequence of one movement with size $2\times 160$. Learning rate for adversarial sample generator is set to $\lambda = 1e-4$.

In the defensive distillation method, the temperature T is set to 10 during training.

\section{Detailed experiment result in adversarial sample set and  normal testing set}
\label{sec:detailed results}

Table.~\ref{table:improved feature selector} and Table.~\ref{table:improved feature selector ts} shows results for the performance of improved feature selector in the adversarial sample set and regular testing set. Table.~\ref{table:basic feature selector} and Table.~\ref{table:basic feature selector ts} shows corresponding results for basic feature selector. We also tested adversarial training based strategy and defensive distillation based strategy for comparison. See Table.~\ref{table:adversarial training} and Table.~\ref{table:defensive distillation} for the results.

\begin{table*}[hbt]
\centering
\caption{True positive rate of improved feature selector based defense strategy in adversarial sample set.}
\label{table:improved feature selector}
\scalebox{0.75}{
\begin{tabular}{|c|c|c|c|c|c|c|c|c|c|c|c|c|c|c|c|c|c|c|c|c|}
\hline
\begin{tabular}[c]{@{}c@{}}True positive\\ rate under\\ adversarial\\ attack\end{tabular} & \begin{tabular}[c]{@{}c@{}}Number of\\ movements\\ been selected\end{tabular} & u1 & u2 & u3 & u4 & u5 & u6 & u7 & u8 & u9 & u10 & u11 & u12 & u13 & u14 & u15 & u16 & u17 & u18 & Mean \\ \hline
\multirow{5}{*}{\begin{tabular}[c]{@{}c@{}}Classifier\\ accessible\end{tabular}} & 10 & 0.068 & 0.035 & 0.13 & 0.035 & 0 & 0.276 & 0.085 & 0 & 0.004 & 0 & 0.004 & 0.087 & 0 & 0 & 0.034 & 0.013 & 0 & 0.087 & 0.048 \\ \cline{2-21} 
 & 2 & 0.817 & 0.667 & 0.492 & 0.987 & 1 & 0.874 & 0.908 & 1 & 0.511 & 0.667 & 0.969 & 1 & 0.719 & 0.307 & 0.015 & 0.962 & 0.143 & 0.996 & 0.724 \\ \cline{2-21} 
 & 3 & 0.909 & 0.667 & 0.408 & 0.948 & 1 & 0.841 & 0.913 & 0 & 0.473 & 1 & 0.328 & 1 & 0.687 & 0.903 & 0.025 & 0.919 & 0.126 & 0.978 & 0.674 \\ \cline{2-21} 
 & 4 & 0.908 & 0.662 & 0.385 & 0.974 & 0.667 & 0.882 & 0.894 & 1 & 0.235 & 0.667 & 0.333 & 1 & 0.835 & 0.307 & 0.024 & 0.97 & 0.134 & 0.965 & 0.658 \\ \cline{2-21} 
 & 5 & 0.566 & 0 & 0.667 & 0.948 & 0.649 & 0.118 & 0.849 & 0 & 0.013 & 1 & 0.313 & 1 & 0.812 & 0.299 & 0.004 & 0.004 & 1 & 0.096 & 0.463 \\ \hline
\multirow{4}{*}{\begin{tabular}[c]{@{}c@{}}Classifier and\\ feature\\ selector\\ accessible\end{tabular}} & 2 & 0.258 & 0.415 & 0.229 & 1 & 0.87 & 0.253 & 0.499 & 0.333 & 0.734 & 0 & 0 & 0.571 & 0.25 & 0 & 0.02 & 0.329 & 0.143 & 1 & 0.384 \\ \cline{2-21} 
 & 3 & 0.272 & 0.088 & 0.144 & 0.758 & 0.341 & 0.212 & 0.585 & 0 & 0.004 & 0.996 & 0 & 0.556 & 0.123 & 0.157 & 0.034 & 0.299 & 0.126 & 1 & 0.316 \\ \cline{2-21} 
 & 4 & 0.395 & 0.108 & 0.177 & 0.587 & 0.035 & 0.301 & 0.497 & 0.333 & 0 & 0.494 & 0 & 0.854 & 0.181 & 0 & 0.02 & 0.342 & 0.134 & 1 & 0.303 \\ \cline{2-21} 
 & 5 & 0.245 & 0.233 & 0.532 & 0.537 & 0.052 & 0.118 & 0.484 & 0 & 0.018 & 0.973 & 0 & 0.569 & 0.162 & 0 & 0.004 & 0.307 & 0.75 & 0.122 & 0.284 \\ \hline
\end{tabular}
}
\vspace{-2mm}
\end{table*}

\begin{table*}[hbt]
\centering
\caption{True positive rate of basic feature selector based defense strategy in adversarial sample set.}
\label{table:basic feature selector}
\scalebox{0.75}{
\begin{tabular}{|c|c|c|c|c|c|c|c|c|c|c|c|c|c|c|c|c|c|c|c|c|}
\hline
\begin{tabular}[c]{@{}c@{}}True positive\\ rate under\\ adversarial\\ attack\end{tabular} & \begin{tabular}[c]{@{}c@{}}Number of\\ movements\\ been selected\end{tabular} & u1 & u2 & u3 & u4 & u5 & u6 & u7 & u8 & u9 & u10 & u11 & u12 & u13 & u14 & u15 & u16 & u17 & u18 & Mean \\ \hline
\multirow{5}{*}{\begin{tabular}[c]{@{}c@{}}Classifier\\ accessible\end{tabular}} & 10 & 0.068 & 0.035 & 0.13 & 0.035 & 0 & 0.276 & 0.085 & 0 & 0.004 & 0 & 0.004 & 0.087 & 0 & 0 & 0.034 & 0.013 & 0 & 0.087 & 0.048 \\ \cline{2-21} 
 & 2 & 0.817 & 0.325 & 0.492 & 0.987 & 1 & 0.874 & 0.908 & 0 & 0.424 & 0.667 & 0.969 & 1 & 0.719 & 0.299 & 0.015 & 0.962 & 0.134 & 0.996 & 0.644 \\ \cline{2-21} 
 & 3 & 0.909 & 0.662 & 0.408 & 0.948 & 1 & 0.841 & 0.913 & 0 & 0 & 1 & 0.328 & 1 & 0.687 & 0.299 & 0.025 & 0.919 & 0.143 & 0.978 & 0.614 \\ \cline{2-21} 
 & 4 & 0.908 & 0.654 & 0.385 & 0.974 & 0.333 & 0.882 & 0.894 & 0.004 & 0.004 & 0.667 & 0.333 & 1 & 0.835 & 0.281 & 0.024 & 0.97 & 0.126 & 0.965 & 0.569 \\ \cline{2-21} 
 & 5 & 0.566 & 0.005 & 0.667 & 0.948 & 0 & 0.118 & 0.849 & 0 & 0.271 & 1 & 0.313 & 1 & 0.812 & 0 & 0.004 & 0.004 & 0.333 & 0.096 & 0.388 \\ \hline
\multirow{4}{*}{\begin{tabular}[c]{@{}c@{}}Classifier and\\ feature\\ selector\\ accessible\end{tabular}} & 2 & 0.258 & 0.039 & 0.229 & 1 & 0.488 & 0.253 & 0.499 & 0 & 0.635 & 0 & 0 & 0.571 & 0.25 & 0 & 0.02 & 0.329 & 0.134 & 1 & 0.317 \\ \cline{2-21} 
 & 3 & 0.272 & 0.135 & 0.144 & 0.758 & 0.607 & 0.212 & 0.585 & 0 & 0.022 & 0.996 & 0 & 0.556 & 0.123 & 0 & 0.034 & 0.299 & 0.143 & 1 & 0.327 \\ \cline{2-21} 
 & 4 & 0.395 & 0.151 & 0.177 & 0.587 & 0 & 0.301 & 0.497 & 0.004 & 0.009 & 0.494 & 0 & 0.854 & 0.181 & 0 & 0.02 & 0.342 & 0.126 & 1 & 0.285 \\ \cline{2-21} 
 & 5 & 0.245 & 0.005 & 0.532 & 0.537 & 0 & 0.118 & 0.484 & 0 & 0.014 & 0.973 & 0 & 0.569 & 0.162 & 0 & 0.004 & 0.307 & 0.333 & 0.122 & 0.245 \\ \hline
\end{tabular}
}
\vspace{-2mm}
\end{table*}

\begin{table*}[hbt]
\centering
\caption{Performance of adversarial training based defense strategy in testing set and adversarial attack sample set.}
\label{table:adversarial training}
\scalebox{0.75}{
\begin{tabular}{|c|c|c|c|c|c|c|c|c|c|c|c|c|c|c|c|c|c|c|c|}
\hline
 & u1 & u2 & u3 & u4 & u5 & u6 & u7 & u8 & u9 & u10 & u11 & u12 & u13 & u14 & u15 & u16 & u17 & u18 & Mean \\ \hline
\begin{tabular}[c]{@{}c@{}}Accuracy in\\ testing set\end{tabular} & 0.882 & 0.863 & 0.788 & 0.911 & 0.917 & 0.894 & 0.974 & 0.966 & 0.873 & 0.788 & 0.833 & 0.882 & 0.821 & 0.923 & 0.833 & 0.833 & 0.981 & 0.807 & 0.876 \\ \hline
\begin{tabular}[c]{@{}c@{}}True positive\\ rate under\\ adversarial\\ attack\end{tabular} & 0 & 0.061 & 0 & 0.056 & 0.139 & 0 & 0.014 & 0.058 & 0 & 0.122 & 0.991 & 0.121 & 0.087 & 0 & 1 & 0 & 0.004 & 0.849 & 0.195 \\ \hline
\end{tabular}
}
\vspace{-2mm}
\end{table*}

\begin{table*}[hbt]
\centering
\caption{Performance of defensive distillation based defense strategy in testing set and adversarial attack sample set.}
\label{table:defensive distillation}
\scalebox{0.75}{
\begin{tabular}{|c|c|c|c|c|c|c|c|c|c|c|c|c|c|c|c|c|c|c|c|}
\hline
 & u1 & u2 & u3 & u4 & u5 & u6 & u7 & u8 & u9 & u10 & u11 & u12 & u13 & u14 & u15 & u16 & u17 & u18 & Mean \\ \hline
\begin{tabular}[c]{@{}c@{}}Accuracy in\\ testing set\end{tabular} & 0.871 & 0.931 & 0.831 & 0.953 & 0.926 & 0.821 & 0.931 & 0.929 & 0.826 & 0.857 & 0.894 & 0.902 & 0.784 & 0.918 & 0.933 & 0.898 & 1 & 0.929 & 0.896 \\ \hline
\begin{tabular}[c]{@{}c@{}}True positive\\ rate under\\ adversarial\\ attack\end{tabular} & 0 & 0 & 0 & 0.265 & 0 & 1 & 0 & 0.052 & 0.214 & 0.346 & 0 & 0.004 & 0 & 0 & 0.005 & 0 & 0.004 & 0.229 & 0.118 \\ \hline
\end{tabular}
}
\vspace{-2mm}
\end{table*}

\begin{table*}[hbt]
\centering
\caption{Accuracy of improved feature selector based defense strategy in testing set.}
\label{table:improved feature selector ts}
\scalebox{0.75}{
\begin{tabular}{|c|c|c|c|c|c|c|c|c|c|c|c|c|c|c|c|c|c|c|c|}
\hline
\begin{tabular}[c]{@{}c@{}}Number of\\ movements\\ been selected\end{tabular} & u1 & u2 & u3 & u4 & u5 & u6 & u7 & u8 & u9 & u10 & u11 & u12 & u13 & u14 & u15 & u16 & u17 & u18 & Mean \\ \hline
10 & 0.923 & 0.911 & 0.894 & 0.947 & 0.962 & 0.965 & 0.953 & 0.918 & 0.833 & 0.863 & 0.91 & 0.903 & 0.841 & 0.922 & 0.934 & 0.882 & 0.966 & 0.913 & 0.913 \\ \hline
2 & 0.613 & 0.655 & 0.615 & 0.641 & 0.685 & 0.732 & 0.621 & 0.804 & 0.681 & 0.653 & 0.809 & 0.647 & 0.549 & 0.918 & 0.733 & 0.729 & 0.878 & 0.679 & 0.702 \\ \hline
3 & 0.661 & 0.828 & 0.708 & 0.781 & 0.685 & 0.857 & 0.721 & 0.893 & 0.71 & 0.674 & 0.851 & 0.647 & 0.49 & 0.837 & 0.717 & 0.797 & 0.918 & 0.749 & 0.751 \\ \hline
4 & 0.71 & 0.862 & 0.677 & 0.828 & 0.815 & 0.911 & 0.855 & 0.857 & 0.826 & 0.714 & 0.872 & 0.784 & 0.608 & 0.898 & 0.8 & 0.831 & 0.898 & 0.816 & 0.809 \\ \hline
5 & 0.855 & 0.897 & 0.677 & 0.859 & 0.852 & 0.946 & 0.972 & 0.839 & 0.812 & 0.735 & 0.915 & 0.843 & 0.667 & 0.898 & 0.867 & 0.881 & 0.918 & 0.886 & 0.851 \\ \hline
\end{tabular}
}
\vspace{-2mm}
\end{table*}

\begin{table*}[hbt]
\centering
\caption{Accuracy of basic feature selector based defense strategy in testing set.}
\label{table:basic feature selector ts}
\scalebox{0.75}{
\begin{tabular}{|c|c|c|c|c|c|c|c|c|c|c|c|c|c|c|c|c|c|c|c|}
\hline
\begin{tabular}[c]{@{}c@{}}Number of\\ movements\\ been selected\end{tabular} & u1 & u2 & u3 & u4 & u5 & u6 & u7 & u8 & u9 & u10 & u11 & u12 & u13 & u14 & u15 & u16 & u17 & u18 & Mean \\ \hline
10 & 0.923 & 0.911 & 0.894 & 0.947 & 0.962 & 0.965 & 0.953 & 0.918 & 0.833 & 0.863 & 0.91 & 0.903 & 0.841 & 0.922 & 0.934 & 0.882 & 0.966 & 0.913 & 0.913 \\ \hline
2 & 0.613 & 0.776 & 0.615 & 0.641 & 0.685 & 0.732 & 0.621 & 0.804 & 0.638 & 0.653 & 0.809 & 0.647 & 0.745 & 0.918 & 0.733 & 0.729 & 0.796 & 0.75 & 0.717 \\ \hline
3 & 0.661 & 0.845 & 0.708 & 0.781 & 0.667 & 0.857 & 0.721 & 0.786 & 0.681 & 0.674 & 0.851 & 0.647 & 0.804 & 0.857 & 0.717 & 0.797 & 0.898 & 0.804 & 0.764 \\ \hline
4 & 0.71 & 0.914 & 0.677 & 0.828 & 0.778 & 0.911 & 0.855 & 0.875 & 0.638 & 0.714 & 0.872 & 0.784 & 0.726 & 0.837 & 0.8 & 0.831 & 0.918 & 0.821 & 0.805 \\ \hline
5 & 0.855 & 0.914 & 0.677 & 0.859 & 0.852 & 0.946 & 0.972 & 0.857 & 0.797 & 0.735 & 0.915 & 0.843 & 0.765 & 0.837 & 0.867 & 0.881 & 0.918 & 0.839 & 0.852 \\ \hline
\end{tabular}
}
\vspace{-2mm}
\end{table*}

%% file: mainfile.bbl
\begin{thebibliography}{10}
\providecommand{\url}[1]{#1}
\csname url@samestyle\endcsname
\providecommand{\newblock}{\relax}
\providecommand{\bibinfo}[2]{#2}
\providecommand{\BIBentrySTDinterwordspacing}{\spaceskip=0pt\relax}
\providecommand{\BIBentryALTinterwordstretchfactor}{4}
\providecommand{\BIBentryALTinterwordspacing}{\spaceskip=\fontdimen2\font plus
\BIBentryALTinterwordstretchfactor\fontdimen3\font minus \fontdimen4\font\relax}
\providecommand{\BIBforeignlanguage}[2]{{%
\expandafter\ifx\csname l@#1\endcsname\relax
\typeout{** WARNING: IEEEtran.bst: No hyphenation pattern has been}%
\typeout{** loaded for the language `#1'. Using the pattern for}%
\typeout{** the default language instead.}%
\else
\language=\csname l@#1\endcsname
\fi
#2}}
\providecommand{\BIBdecl}{\relax}
\BIBdecl

\bibitem{s20226578}
\BIBentryALTinterwordspacing
I.~Vaccari, G.~Chiola, M.~Aiello, M.~Mongelli, and E.~Cambiaso, ``Mqttset, a new dataset for machine learning techniques on mqtt,'' \emph{Sensors}, vol.~20, no.~22, 2020. [Online]. Available: \url{https://www.mdpi.com/1424-8220/20/22/6578}
\BIBentrySTDinterwordspacing

\bibitem{2015DNS}
M.~Aiello, M.~Mongelli, and G.~Papaleo, ``Dns tunneling detection through statistical fingerprints of protocol messages and machine learning,'' \emph{International Journal of Communication Systems}, vol.~28, no.~14, 2015.

\bibitem{2020Adversarial}
W.~Jin, Y.~Li, H.~Xu, Y.~Wang, and J.~Tang, ``Adversarial attacks and defenses on graphs: A review and empirical study,'' 2020.

\bibitem{2021Adversarial}
Y.~Pacheco and W.~Sun, ``Adversarial machine learning: A comparative study on contemporary intrusion detection datasets,'' in \emph{7th International Conference on Information Systems Security and Privacy}, 2021.

\bibitem{0The}
S.~Sabour, Y.~Cao, F.~Faghri, and D.~J. Fleet, ``The limitations of deep learning in adversarial settings.''

\bibitem{DBLP:journals/corr/abs-1811-01439}
\BIBentryALTinterwordspacing
B.~D. Mittelstadt, C.~Russell, and S.~Wachter, ``Explaining explanations in {AI},'' \emph{CoRR}, vol. abs/1811.01439, 2018. [Online]. Available: \url{http://arxiv.org/abs/1811.01439}
\BIBentrySTDinterwordspacing

\bibitem{lipton2018mythos}
Z.~C. Lipton, ``The mythos of model interpretability: In machine learning, the concept of interpretability is both important and slippery.'' \emph{Queue}, vol.~16, no.~3, pp. 31--57, 2018.

\bibitem{silver2017mastering}
D.~Silver, J.~Schrittwieser, K.~Simonyan, I.~Antonoglou, A.~Huang, A.~Guez, T.~Hubert, L.~Baker, M.~Lai, A.~Bolton \emph{et~al.}, ``Mastering the game of go without human knowledge,'' \emph{nature}, vol. 550, no. 7676, pp. 354--359, 2017.

\bibitem{du2019techniques}
M.~Du, N.~Liu, and X.~Hu, ``Techniques for interpretable machine learning,'' \emph{Communications of the ACM}, vol.~63, no.~1, pp. 68--77, 2019.

\bibitem{dabkowski2017real}
P.~Dabkowski and Y.~Gal, ``Real time image saliency for black box classifiers,'' \emph{Advances in neural information processing systems}, vol.~30, 2017.

\bibitem{chen2018learning}
J.~Chen, L.~Song, M.~Wainwright, and M.~Jordan, ``Learning to explain: An information-theoretic perspective on model interpretation,'' in \emph{International Conference on Machine Learning}.\hskip 1em plus 0.5em minus 0.4em\relax PMLR, 2018, pp. 883--892.

\bibitem{yoon2018invase}
J.~Yoon, J.~Jordon, and M.~van~der Schaar, ``Invase: Instance-wise variable selection using neural networks,'' in \emph{International Conference on Learning Representations}, 2018.

\bibitem{fu2021differentiated}
W.~Fu, M.~Wang, M.~Du, N.~Liu, S.~Hao, and X.~Hu, ``Differentiated explanation of deep neural networks with skewed distributions,'' \emph{IEEE Transactions on Pattern Analysis and Machine Intelligence}, 2021.

\bibitem{mathew2019handedness}
J.~Mathew, F.~R. Sarlegna, P.-M. Bernier, and F.~R. Danion, ``Handedness matters for motor control but not for prediction,'' \emph{eneuro}, vol.~6, no.~3, 2019.

\bibitem{QIN2024106166}
Qin, D., Amariucai, G.~T., Qiao, D., Guan, Y., and Fu, S. (2024).
\newblock A comprehensive and reliable feature attribution method: Double-sided remove and reconstruct (DoRaR).
\newblock {\em Neural Networks}, 173:106166.

\bibitem{fu2022artificial}
S.~Fu, D.~Qin, G.~Amariucai, D.~Qiao, Y.~Guan, and A.~Smiley, ``Artificial intelligence meets kinesthetic intelligence: Mouse-based user authentication based on hybrid human-machine learning,'' in \emph{Proceedings of the 2022 ACM on Asia Conference on Computer and Communications Security}, 2022, pp. 1034--1048.

\bibitem{DBLP:journals/corr/GoodfellowSS14}
\BIBentryALTinterwordspacing
I.~J. Goodfellow, J.~Shlens, and C.~Szegedy, ``Explaining and harnessing adversarial examples,'' in \emph{3rd International Conference on Learning Representations, {ICLR} 2015, San Diego, CA, USA, May 7-9, 2015, Conference Track Proceedings}, Y.~Bengio and Y.~LeCun, Eds., 2015. [Online]. Available: \url{http://arxiv.org/abs/1412.6572}
\BIBentrySTDinterwordspacing

\bibitem{7780651}
S.-M. Moosavi-Dezfooli, A.~Fawzi, and P.~Frossard, ``Deepfool: A simple and accurate method to fool deep neural networks,'' in \emph{2016 IEEE Conference on Computer Vision and Pattern Recognition (CVPR)}, 2016, pp. 2574--2582.

\bibitem{2013Intriguing}
C.~Szegedy, W.~Zaremba, I.~Sutskever, J.~Bruna, D.~Erhan, I.~Goodfellow, and R.~Fergus, ``Intriguing properties of neural networks,'' \emph{Computer Science}, 2013.

\bibitem{2016Distillation}
N.~Papernot, P.~Mcdaniel, X.~Wu, S.~Jha, and A.~Swami, ``Distillation as a defense to adversarial perturbations against deep neural networks,'' in \emph{2016 IEEE Symposium on Security and Privacy (SP)}, 2016.

\bibitem{2015Distilling}
G.~Hinton, O.~Vinyals, and J.~Dean, ``Distilling the knowledge in a neural network,'' \emph{Computer Science}, vol.~14, no.~7, pp. 38--39, 2015.

\bibitem{2016Towards}
N.~Carlini and D.~Wagner, ``Towards evaluating the robustness of neural networks,'' 2016.

\bibitem{2009Evade}
B.~Biggio, G.~Fumera, and F.~Roli, ``Evade hard multiple classifier systems,'' \emph{Studies in Computational Intelligence}, 2009.

\bibitem{Battista2013Security}
Battista, Biggio, Giorgio, Fumera, Fabio, and Roli, ``Security evaluation of pattern classifiers under attack,'' \emph{IEEE Transactions on Knowledge and Data Engineering}, 2013.

\bibitem{2014Feature}
B.~Li and Y.~Vorobeychik, ``Feature cross-substitution in adversarial classification,'' in \emph{Neural Information Processing Systems}, 2014.

\bibitem{Fei2016Adversarial}
Fei, Zhang, Patrick, P, K, Chan, Battista, Biggio, Daniel, and S, ``Adversarial feature selection against evasion attacks.'' \emph{IEEE Transactions on Cybernetics}, 2016.

\bibitem{8249883}
Z.~Yin, F.~Wang, W.~Liu, and S.~Chawla, ``Sparse feature attacks in adversarial learning,'' \emph{IEEE Transactions on Knowledge and Data Engineering}, vol.~30, no.~6, pp. 1164--1177, 2018.

\end{thebibliography}
